\documentclass[12pt,preprint]{aastex}
\usepackage{bbm}
\usepackage{rotating}
\usepackage[norefs,nocites]{refcheck}
\usepackage{amsmath}

\newcommand{\beq}{\begin{eqnarray}}
\newcommand{\eeq}{\end{eqnarray}}
\shorttitle{Dark Energy from SNe Ia and Strong Gravitational Lenses}
\shortauthors{Yue-Liang Wu, et al.}

\begin{document}

\title{  Time-Varying Dark Energy Constraints From the Latest SN Ia, BAO and SGL}
\author{Qing-Jun Zhang$^{1,2}$ and Yue-Liang Wu$^{1}$}

\affil{ $^1$ \rm  Kavli Institute for Theoretical Physics China,
\\ Key Laboratory of Frontiers in  Theoretical Physics, \\ Institute of
Theoretical Physics, Chinese Academy of Science, Beijing 100190,
China}

\affil{ $^2$ \rm Department of Physics, Graduate University,
\\the Chinese Academy of Sciences YuQuan Road 19A, 100049, Beijing,
China}

 \email{ylwu@itp.ac.cn}

\begin{abstract}
Based on the latest SNe Ia data provided by Hicken et al. (2009)
with using MLCS17 light curve fitter, together with the Baryon
Acoustic Oscillation(BAO) and strong gravitational lenses(SGL), we
investigate the constraints on the dark energy equation-of-state
parameter $w$ in the flat universe, especially for the time-varying
case $w(z)=w_0+w_zz/(1+z)$. The constraints from SNe data alone are
found to be: (a) $(\Omega_M, w)=(0.358, -1.09)$ as the best-fit
results; (b) $(w_0, w_z)=(-0.73^{+0.23}_{-0.97},
0.84^{+1.66}_{-10.34})$  for the two parameters in the time-varying
case  after marginalizing the parameter $\Omega_M$; (c) the
likelihood of parameter $w_z$ has a high non-Gaussian distribution;
(d) an extra restriction on $\Omega_M$ is necessary to improve the
constraint of the SNe Ia data on the parameters ($w_0$, $w_z$). A
joint analysis of SNe Ia data and BAO is made to break the
degeneracy between $w$ and $\Omega_M$, and leads to the interesting
maximum likelihoods $w_0 = -0.94$ and $w_z = 0$. When marginalizing
the parameter $\Omega_M$, the fitting results are found to be $(w_0,
w_z)=(-0.95^{+0.45}_{-0.18}, 0.41^{+0.79}_{-0.96})$. After adding
the splitting angle statistic of SGL data, a consistent constraint
is obtained $(\Omega_M, w)=(0.298, -0.907)$ and the constraints on
time-varying dark energy are further improved to be $(w_0, w_z) =
(-0.92^{+0.14}_{-0.10}, 0.35^{+0.47}_{-0.54})$, which indicates that
the phantom type models are disfavored.
\end{abstract}





\section{INTRODUCTION}

  Analysis of the distance modulus versus redshift relation of type
Ia supernova (SNe Ia) provides a direct evidence that the universe
expansion is accelerating in the last few billion years(e.g.,
Perlmutter et al. 1999, Riess et al. 1998, 2004, 2007; Astier et al.
2006; Wood-Vasey et al. 2007; Kowalski et al. 2008; Hicken et al.
2009;  Kessler et al. 2009). This cosmic image is also supported by
many other cosmological observations, like the Cosmic Microwave
Background(CMB)(Hinshaw et al. 2009; Komatsu et al. 2009), the
Baryon Acoustic Oscillation (BAO) measurement(Eisenstein et al.
2005; Cole et al. 2005; Huetsi 2006; Percival et al. 2007) and the
weak gravitational lenses (Weinberg and Kamionkowski 2002; Zhan and
Knox 2006). Based on the Friedmann equation, the acceleration can be
explained through introducing a negative pressure component in the
universe, named dark energy, which is nearly spatially uniform
distribution and contributes about $2/3$ critical density of
universe today. To reveal the property of dark energy, the most of
studies, either theoretical models or experiment data analysis, are
focused on its equation-of-state parameter $w=p/\rho$. Here we shall
utilize the latest SNe Ia data provided by Hicken et al. (2009) with
using MLCS17 light curve fitter, together with the splitting angle
statistic of strong gravitational lenses(SGL; Zhang et al. 2009) and
the baryonic acoustic oscillations (BAO; Eisenstein et al. 2005) to
investigate the constraint for the parameter $w$ of dark energy in
the flat cosmology, especially for the time varying case.

 By far, all observed data are consistent with the $\Lambda {\rm
CDM}$ cosmology, with dark energy in the form of a cosmological
constant $\Lambda$. However, this model raises theoretical problems
related to the fine tuned value (see e.g. Padmanabhan 2003). Many
other theoretical models, like quintessence models and phantom
model(Ratra and Peebles 1988; Caldwell, Dav{\'e}, and Steinhardt
1998; Caldwell 2002) , reveal that dark energy might be a dynamical
component and evolves with time. It is usual to parametrize dark
energy as an ideal liquid  with its equation-of-state(EOS) parameter
$w(z)=w_0 + w_z \; z/(1+z)$ (Chevallier et al. 2001; Linder 2003).
Conveniently, it includes the case of a constant EOS with ($w_0 =
w,w_z = 0$), and the $\Lambda {\rm CDM}$ model ($w_0 = -1,w_z = 0$).
Then theoretical models can be classified in a phase diagram on the
($w_0,w_z$) plane (see e.g. Barger et al. 2006; Biswas and Wandelt
2009). Thus the accurate measurement of the parameters ($w_0, w_z$)
is helpful for testing a certain theoretical models. The current
allowed regions of ($w_0, w_z$) given by the most observations or
their combinations remain surrounding the crucial point ($w_0 =-1,
w_z=0$), which is the common point in the phase diagram of different
classified models. Therefore, the final judgment of models can not
be made and more careful works are still needed.

The SNe Ia has homogeneity and extremely high intrinsic luminosity
of peak magnitude and thus is widely used to measure the
cosmological parameters(e.g. Riess et al. 2004; Wood-Vasey et al.
2007; Kowalski et al. 2008; Hicken et al. 2009; Kessler et al.
2009). With given density of dark energy in the universe today
$\rho(z=0)$, the change of equation-of-state parameter $w$ will
bring the change of its density $\rho(z)$ at the redshift $z$ and
then the distance $d(z)$. Inversely, the measurements of the
redshift $z$  of SNe Ia and corresponding distance $d(z)$ can
constrain the dark energy. In spite of the high accuracy of SNe Ia
measurement, its potential of constraining dark energy is not very
strong due to the degeneracy of $w$ and $\Omega_M$. Therefore, a
combining analysis with other observations is often useful.

We shall also use the  summary parameters of the baryonic acoustic
oscillations as reported in previous studies(Eisenstein et al.
2005). The large-scale correlation function of a large sample of
luminous red galaxies has been measured in the Sloan Digital Sky
Survey and a well-detected acoustic peak was found to provide a
standard ruler by which the absolute distance of $z=0.35$ can be
determined with $5\%$ accuracy,  which is independent of the Hubble
constant $h$. This ruler is a $\Omega_M$ prior and can be used to
constrain dark energy(e.g. Porciani and Madau 2000; Huterer and Ma
2004; Chae, 2007). We are going to show that the strong
gravitational lensing statistic observation can also provide us a
useful probe of dark energy of the universe. This is because the
dark energy affects, mainly through the comoving number density of
dark halos described by Press-Schechter theory and the background
cosmological line element,  the efficiency with which dark-matter
concentrations produce strong lensing signals. Then by comparing the
observed number of lenses with the theoretical expected result as a
function of image separation and cosmological parameters, it enables
us to determine the allowed range of the parameter $w$. The
constraint process also depends on the density profile of dark
halos. Here we will use the two model combined mechanism to
reproduce the observed curve of lensing probability to the image
splitting angle (Sarbu, Rusin and Ma 2001; Li and Ostriker 2002;
Zhang et al. 2009). The redshift of ${\rm CMB}$ is above $1000$ and
far larger than $1$, and there is no other observation to fill up
this redshift gap, thus we would not adopt the CMB data in the
present analysis and limit our study on dark energy to the redshift
region of $z \sim 1$, which is the characteristic redshift scale of
SNe Ia, BAO and SGL statistic.

In our recent work(Zhang et al. 2009), we have present the
constraint on the dark energy from the SGL splitting angle
statistic. In this paper, by taking the latest analyzed SNe Ia
data(Hicken et al. 2009), the baryonic acoustic
oscillations(Eisenstein et al. 2005) and the CLASS statistical
sample(Browne et al.2003), we shall make a joint analysis to
constrain the dark energy equation of state parameters $w$,
especially for the time-varying parameterization $w(z) = w_0 + w_z
\, z / (1 + z)$. We mainly highlight two issues which have not
previously been illuminated. First, we carefully study based on the
latest SNe Ia data the constraints for the dark energy EOS $w(z)$
and the influences of $\Omega_M$. Second, we investigate the joint
analysis of SNe Ia data, BAO and SGL statistic in detail. Our paper
is organized as follows: Sect. 2 shows the constraint by the latest
SNe Ia data on dark energy. Sect. 3 describes the joint analysis of
the SNe Ia, the BAO and the SGL statistic, more stringent
constraints on dark energy are resulted. The conclusions are
presented in the last section.

\section{DARK ENERGY CONSTRAINTS BY THE LATEST SNe Ia DATA}

    As the standard candles of the cosmology, the SNe Ia is used to
study the geometry and dynamics of the universe with redshift $z\leq
1.7$. In determinations of cosmological parameters about the
accelerating expand and dark energy, the SNe Ia remains a key
ingredient. In 1998, the SNe Ia measurement provided the first
direct evidence for the presence of dark energy with the negative
pressure. Then many SN Ia observations have been done and the total
number of SNe Ia sample increases quickly. The SN Ia compilations
are often consist of high-redshift $(z \simeq 0.5)$ data set and
low-redshift $(z \simeq 0.05)$ sample at the same time (e.g. Riess
et al. 1998; Perlmutter et al. 1999; Wood-Vasey et al. 2007). When
combining several independent group's SNe Ia data sets into one
compilation, the consistent  analysis method of light curves and the
selection of supernova are crucial. For a certain sample, the
different light curve fitter and corresponding different selection
of supernova can lead to different constraints on the  cosmological
parameters(e.g. Hicken et al. 2009).

The fitting results of cosmological parameters from different SN Ia
compilations have moderate differences. To obtain the consistent and
more powerful constraint, researchers have made many efforts to deal
with the cross-calibration uncertainities when combining the
different SNe samples. There is a conventional method to combine
several group's SNe Ia compilations, namely, by introducing an extra
nuisance parameter in the $\chi^2$ statistic of every used SNe Ia
sample and marginalizing them over in the fit, all $\chi^2$
statistics of samples can be summed into one total statistics (see
e.g. Barger et al. 2006). The nuisance parameters are considered as
analysis-dependent global unknown constants in the distances.
Although this combined mechanism is wildly adopted, the so-called
analysis-dependent unknown constant is just an averaged effect of
analysis-dependent uncertainties.

Kowalski et al. (2008) provided the Union data set, a compilation of
307 SNe Ia discovered in different surveys. The heterogeneous nature
of the data set have been reflected and all SNe Ia sample are
analyzed with the same analysis procedure. In the Union data set,
all SNe Ia light curve are fitted by using the
spectral-template-based fit method of Guy et al. (2005) (also known
as SALT). There are other light curve fitters used in literatures,
such as SALT2 (Guy et al. 2007), MLCS2k2 (Jha, Riess, and Kirshner
2007) with $R_V = 3.1$ (MLCS31) and MLCS2k2 with $R_V = 1.7$
(MLCS17). Hicken et al. (2009) compared these light curve fitters
and found that SALT produces high-redshift Hubble residuals with
systematic trends versus color and larger scatter than MLCS2k2, and
MLCS31 overestimates host-galaxy extinction while MLCS17 does not.
For a certain SNe Ia, the analysis outcomes of different light curve
fitters are not equal. Here we choose the SNe Ia compilation
provided by using MLCS17 light curve fitter with the best cuts $A_V
\leq 0.5$ and $\Delta < 0.7$ to constrain the dark energy.

In the flat universe, the Friedmann equation are given by
 \beq
 H(z)/H_0 &=&  \sqrt{ \Omega_M (1+z)^3 +
 \Omega_{DE}(z)}
 \nonumber \\
\Omega_{DE}(z) &=& \left\{\begin{array}{ll}(1 - \Omega_M) (1
+z)^{3(1+w)} & for \;\;  constant \; w \,,\\
            (1 - \Omega_M) (1
+z)^{3(1+w_0+w_z)}e^{-3w_zz/(1+z)}
        \hspace{0.5cm} & for \;\; w(z)=w_0 + w_z \, {z\over1+z} \,,\\
   \end{array} \label{eq:hub}
   \right.
 \eeq with Hubble constant $H_0 = 100 \; h \;{\rm km \;s^{-1}\;
 Mpc^{-1}}$.  The influence of cosmological parameter $w$ is focused on the
 dark energy density $\Omega_{DE}(z)$ and then the Luminosity distance
$d_L$, which is defined as
 \beq
 d_L(z) = (1+z) \int^z_0 {dz' \over H(z')} \label{eq:lumdis}
 \eeq

Analysis of the distance modulus versus redshift relation of SNe Ia
can give us the information about the cosmological parameters.
Distance estimates of SNe Ia are derived from the luminosity
distance, $d_{L} = ({{\cal L} / 4 \pi {\cal F}})^{1/2}$ where ${\cal
L}$ and ${\cal F}$ are the intrinsic luminosity and observed flux of
the SNe Ia, respectively. It is usual to introduce the apparent
magnitude $m$  and absolute magnitude $M$. From the definition of
the distance moduli $\mu=m-M$, we have
 \beq
 \mu=5\log d_L/{\rm Mpc}+25. \label{eq:dismod}
 \eeq
Using Equations (\ref{eq:hub}), (\ref{eq:lumdis}) and
(\ref{eq:dismod}), we can relate the parameter $w$ with the measured
redshift $z$ and distance moduli $\mu(z)$ of SNe Ia data. Since
parameter $H_0$ is irrelevant for the SN only data, the likelihood
of the SNe Ia analysis can be determined from a $\chi^2$ statistic
 \beq
 \chi^2(\Omega_M,w) = \sum_i { (\mu^T_i(z_i;\Omega_M, w)-\mu^O_i)^2\over \sigma_i^2}
 \label{eq:snechi2}
 \eeq
where subscript $i$ denotes the $i$th SNe Ia data and $\sigma_i$ is
the observed uncertainty. $\mu^O$ and $\mu^T$ are the observed and
theoretical distance moduli, respectively.

\begin{figure}
\epsscale{0.8} \plotone{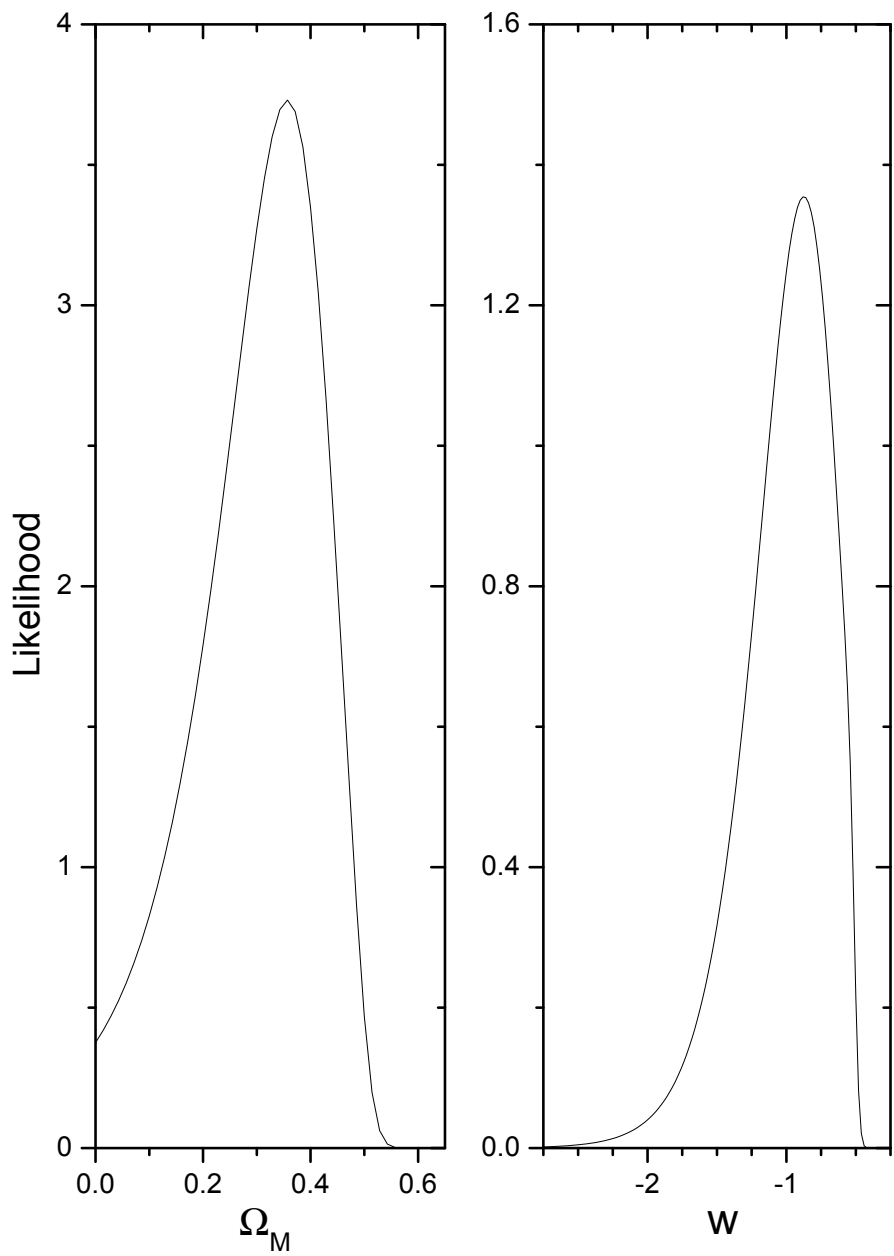}
\figcaption{\small The likelihoods of the parameters $\Omega_M$ and
$w$. The maximum likelihood points are located at $\Omega_M=0.36$,
 and $w=-0.88$, respectively. \label{fig:SNe1}}
\end{figure}

Let us first discuss the constraints for the constant $w$ case.
Using the Powell minimization method(Press, et al 1992), we minimize
the likelihood function of the parameters $(\Omega_M, w)$ and find
that the coordinate of the best-fit point is $(\Omega_M, w) =
(0.358, -1.09)$. Though the results of SNe Ia data and WMAP
observation are consistent in the statistical meaning, it is noted
that this best-fit value of $\Omega_M$ is large, in comparison with
$\Omega_M = 0.258$ of the concordance cosmology provided by WMAP
five year data(Komatsu et al. 2009). Figure \ref{fig:SNe1} shows the
likelihoods of parameter $\Omega_M$ and $w$, in which the maximum
likelihood points are located at $\Omega_M=0.36$ and $w=-0.88$,
respectively. It is interesting to notice that the parameter $w$ is
restricted to be from $-2.0$ to $-0.5$ and $\Omega_M$ is less than
$0.5$.

\begin{figure}
\epsscale{0.8} \plotone{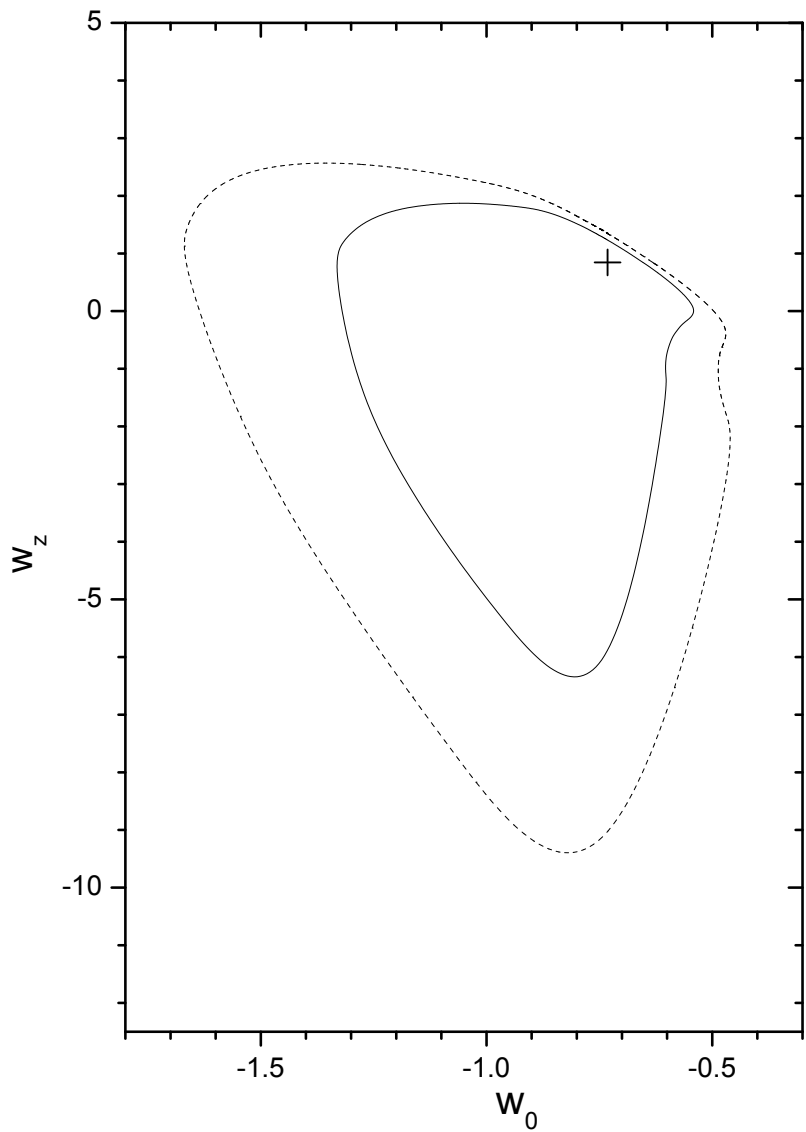}
\figcaption{\small The ($w_0, w_z$) contours of SNe Ia data alone.
The crosshairs mark the best-fit point ($w_0, w_z)=(-0.73,0.84)$.
\label{fig:SNe2} }
\end{figure}

\begin{figure}
\epsscale{0.8} \plotone{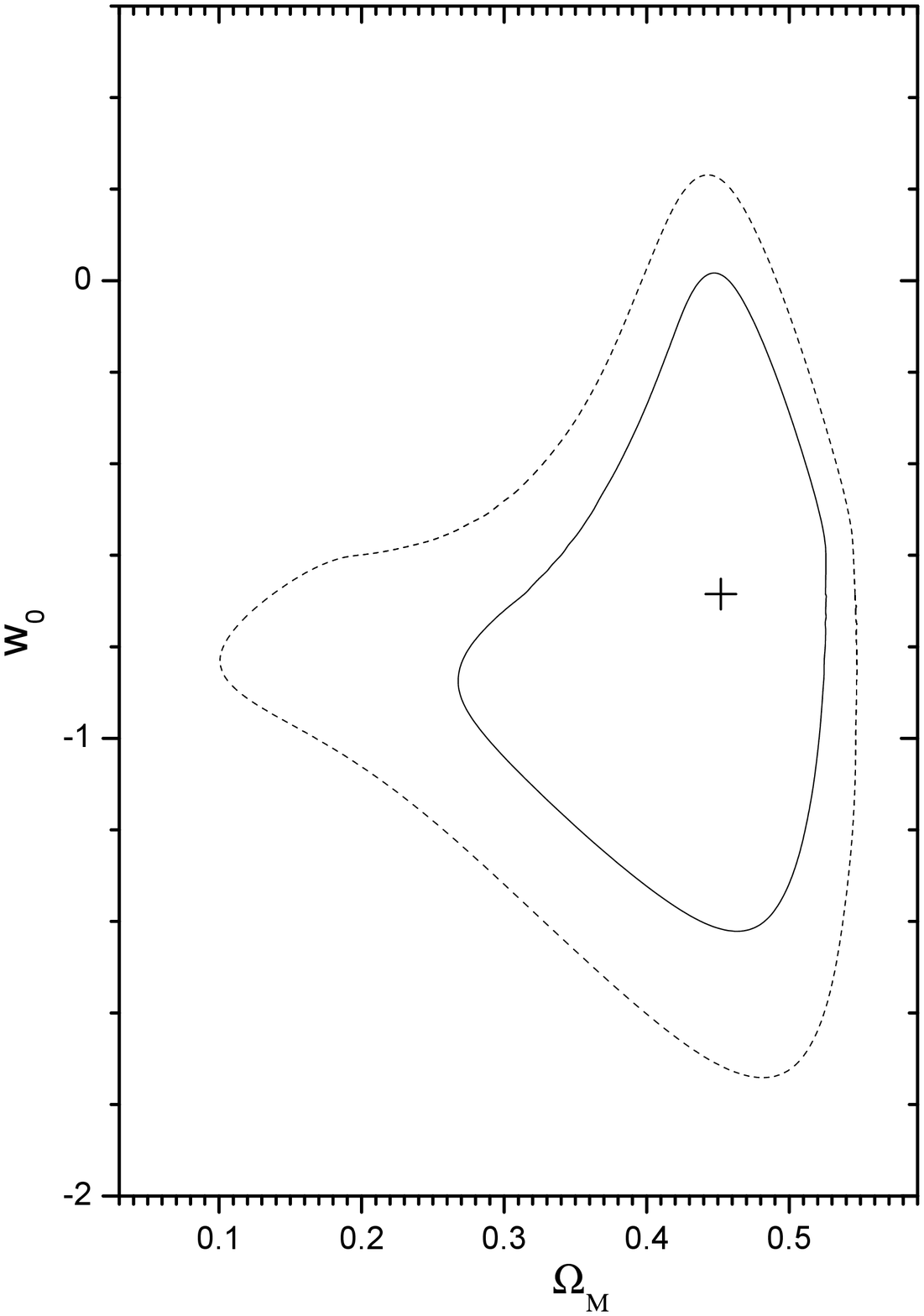}
\figcaption{\small The ($\Omega_M, w_0$) contours of SNe Ia data
alone  after marginalizing $w_z$. The crosshairs mark the best-fit
point ($\Omega_M, w_0)=(0.45, -0.68)$.  \label{fig:SNe3} }
\end{figure}

\begin{figure}
\epsscale{0.8} \plotone{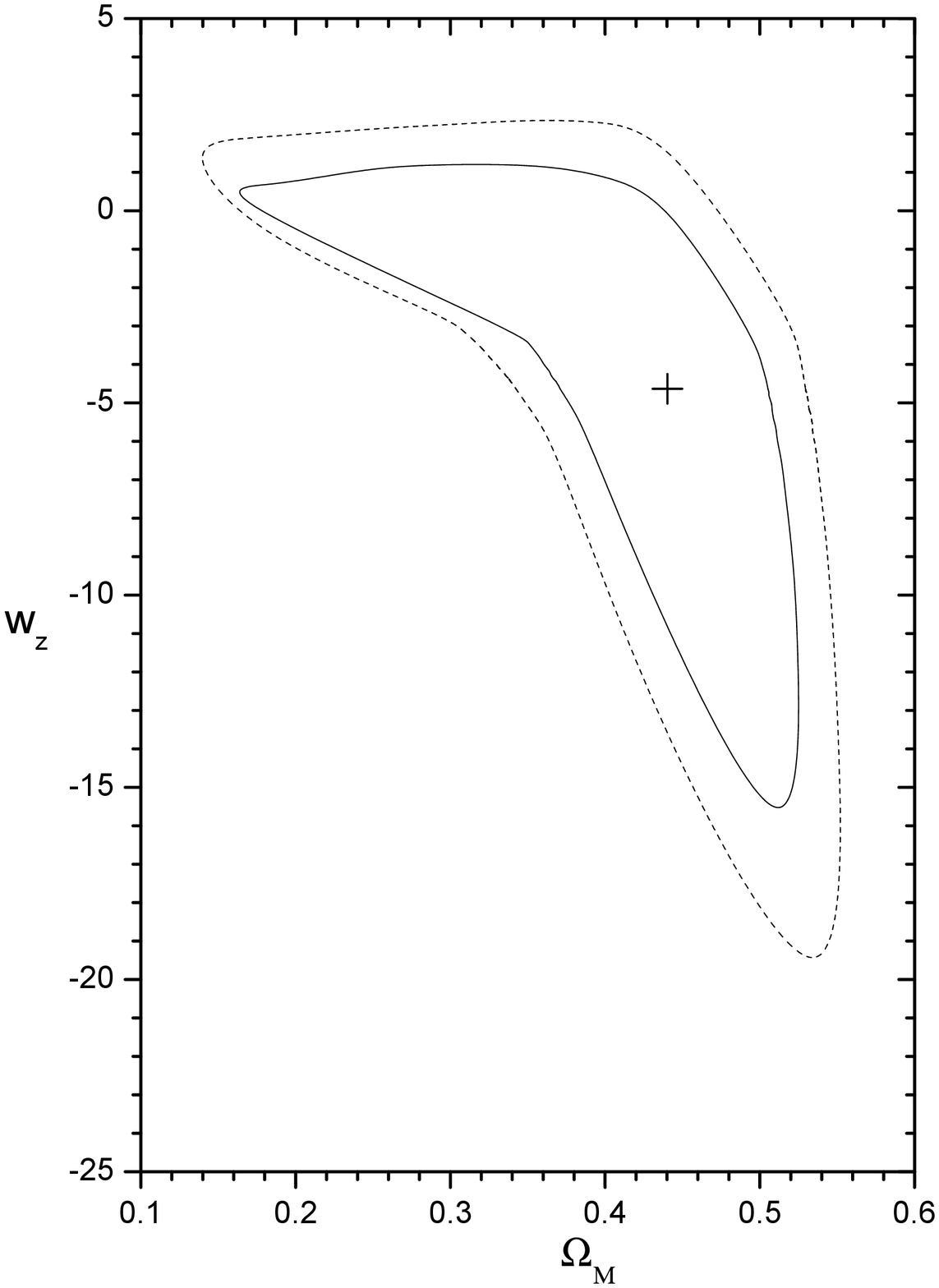}
\figcaption{\small The ($\Omega_M, w_z$) contours of SNe Ia data
alone  after marginalizing $w_0$. The crosshairs mark the best-fit
point ($\Omega_M, w_z)=(0.44, -4.63)$. \label{fig:SNe4}}
 \end{figure}

\begin{figure}
\epsscale{0.8} \plotone{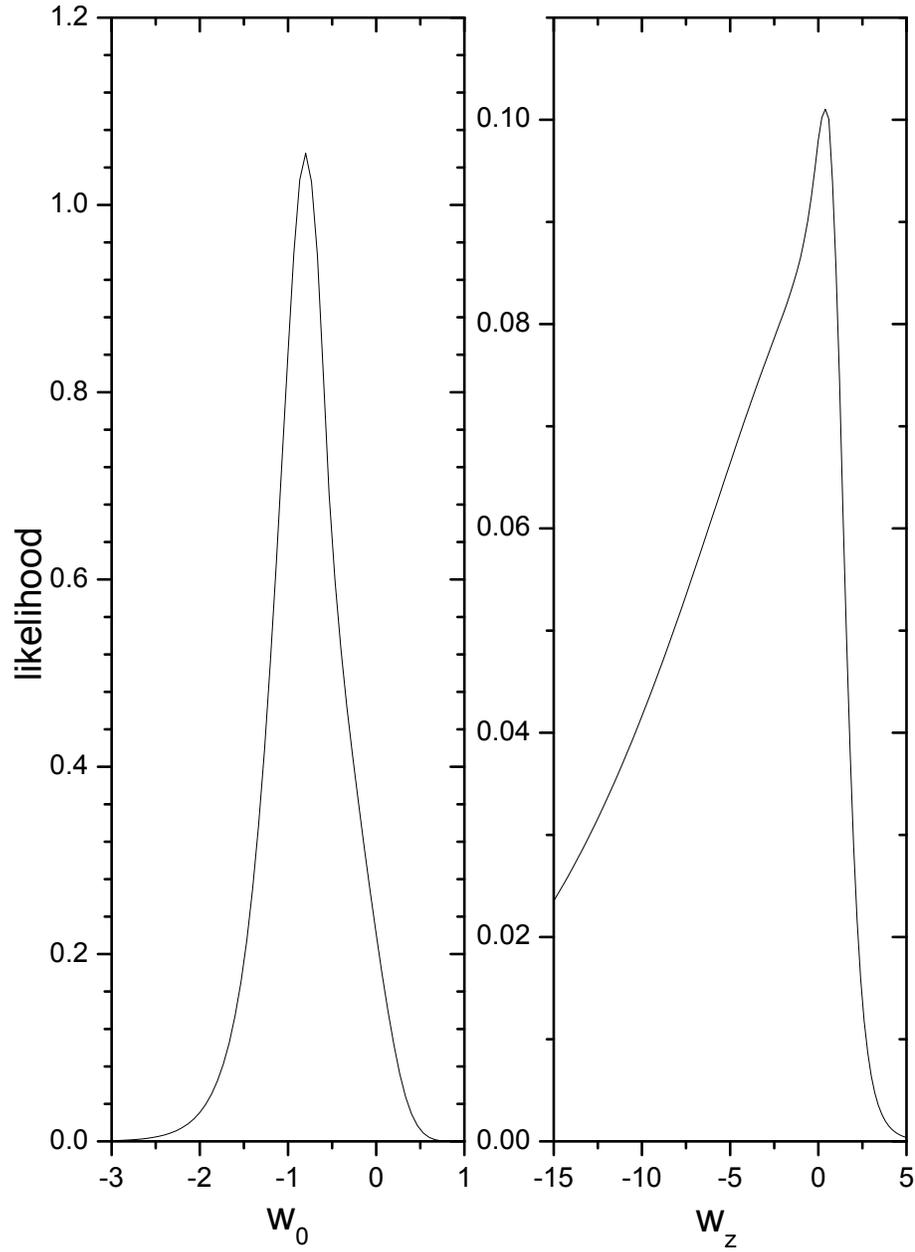}
\figcaption{\small The likelihoods of the parameters $w_0$ and
$w_z$. The maximum likelihood points are located at $w_0=-0.8$ and $
w_z = 0.4$, respectively. It can be seen that the likelihood of
parameter $w_z$ has a high non-Gaussian distribution.
\label{fig:SNe5} }
\end{figure}

We now focus on the time-varying model $w(z) = w_0 + w_z \, z /
(1+z)$. Figure \ref{fig:SNe2} shows the contours of ($w_0, w_z$),
the best-fit point is ($w_0, w_z)=(-0.73,0.84)$ after marginalizing
the parameter $\Omega_M$. It is seen that the SNe Ia data alone have
a poor constraint power on the parameter $w_z$. In figure
\ref{fig:SNe3}, we plot the contours of two parameters ($\Omega_M,
w_0$) after marginalizing $w_z$, the best-fit point is found to be
($\Omega_M, w_0)=(0.45,-0.68)$. It is shown that when $\Omega_M$
increases from $0.3$ to $0.45$, the allowed region of parameter
$w_0$ is enlarged quickly. For a smaller $\Omega_M < 0.3$, it leads
to  a much better constraint for the parameter $w_0$: $w_0 \sim -1.4
\sim -0.6$. Figure \ref{fig:SNe4} gives the contours of ($\Omega_M,
w_z$) after marginalizing $w_0$, the best-fit point is ($\Omega_M,
w_z)=(0.44,-4.63)$. It is noticed that when $\Omega_M$ increases
from $0.34$ to $0.5$, the allowed region for the parameter $w_z$ is
enlarged rapidly. For a smaller $\Omega_M < 0.34$, it also leads to
a much better constraint for the parameter $w_z$: $w_z \sim - 3.0
\sim 2.5$. From the figure \ref{fig:SNe3} and figure \ref{fig:SNe4},
it indicates that  an extra restriction on $\Omega_M$ is necessary
to improve the constraint of the SNe Ia data on the parameters $w_0$
and $w_z$. Figure \ref{fig:SNe5} shows the likelihoods of parameters
$w_0$ and $w_z$, in which the maximum likelihood points are located
at $w_0=-0.8$ and $ w_z = 0.4$, respectively. It can be seen that
the parameter $w_0$ is limited in the region $-2.5 <w_0 <0.5$ and
the likelihood of parameter $w_z$ has a high non-Gaussian
distribution.

\section{JOINT ANALYSIS OF SNe Ia DATA, BAO AND SGL STATISTIC}

As we  have shown in the previous chapter, over $300$ SNe Ia
observed so far are not sufficient for determining the cosmological
parameters, especially for $w_0$ and $w_z$. Many surveys(e.g. the
Dark Energy Survey and Pan-STARRS) are proposed to obtain the SNe Ia
sample with enlarged number and improved precision. Here we are
going to constrain the dark energy through the combination of the
SNe Ia data, the baryon acoustic oscillations as well as the  SGL
splitting angel statistic.

\subsection{Baryon Acoustic Oscillations}

In the relativistic plasma of the early universe, ionized hydrogens
(protons and electrons) are coupled with energetic photons by
Thomson scattering. The plasma density is  uniform except for the
primordial cosmological perturbations. Driven by high pressure, the
plasma fluctuations spread outward at over half the speed of light.
After about $10^5$ years, the universe has cooled enough and the
protons capture the electrons to form neutral Hydrogen. This
decouples the photons from the baryons, which dramatically decreases
the sound speed  and effectively ends the sound wave propagation.
Because the universe has a significant fraction of baryons, these
baryon acoustic oscillations leave their imprint on very large scale
structures (about $100 {\rm Mpc}$) of the Universe.

The measurement of baryon acoustic oscillations was first processed
by the Sloan Digital Sky Survey (SDSS; York et al. 2000) and
Eisenstein et al. (2005) studied the large-scale correlation
function of its sample, which is composed of $46,748$ luminous red
galaxies over 3816 square degrees and in the redshift range $0.16$
to $0.47$. The typical redshift of the sample is at $z = 0.35$. The
large-scale correlation function is a combination of the
correlations measured in the radial (redshift space) and the
transverse (angular space) direction (Davis et al. 2007). Thus, the
relevant distance measure is modeled by the so-called dilation
scale, $D_V(z) = [D^2_A(z) z / H(z)]^{1/3}$, with comoving angular
diameter distance $D_A(z) = \int_0^z dz'/H(z')$. The dimensionless
combination $A(z) = D_V (z) \sqrt{\Omega_M H_0^2}/z$ has no
dependence on the Hubble constant $h$ and is found to be well
constrained by the SDSS data at $z=0.35$. A standard ruler is
provided as (Eisenstein et al. 2005)
 \beq
A = \frac{\sqrt{\Omega_M}}{[H(z_1)/H_0]^{1/3}}~\bigg\lbrack
~\frac{1}{z_1}~\int_0^{z_1}\frac{ dz}{H(z)/H_0}
~\bigg\rbrack^{2/3} = ~0.469  \pm 0.017~, \eeq where $z_1 = 0.35$.
This ruler is a $\Omega_M$ prior and can be used to constrain dark
energy. Then the statistic is given by
 \beq
 \chi^2(\Omega_M, w) = {[A(\Omega_M, w) - 0.469]^2 \over 0.017^2 }
 \label{eq:baochi2}
 \eeq

\subsection{SGL Splitting Angle Statistic}

The CLASS statistical sample has provided a well-defined statistical
sample with  $N=8958$ sources. Totally  $N_l=13$
 multiple image gravitational lenses have been discovered and all have image
separations $\Delta\theta<3^{\prime\prime}$ (Browne et al. 2003).
The SGL statistics are sensitive to the equation-of-state
parameter $w$ of dark energy, which influences the number density
of lens galaxies and the distances between the sources and lens.
The probability with image separations larger than $\Delta\theta$
for a source at redshift $z_s$ on account of the galaxies
distribution from the source to the observer can be obtained by
(Schneider et al. 1992)
 \beq
 P(>\Delta\theta) = \int_0^{z_s}\int_0^{\infty} {d D_L\over dz}
 (1+z)^3 n(M,z)\sigma(>\Delta\theta)\ dM dz\,,
 \label{eq:lp} \eeq
where $M$ is the mass of a dark halo, $D_L$ is the proper distance
from the observer to the lens, n(M, z) is the comoving number
density of dark halos virialized by redshift $z$ with mass $M$ and
$\sigma$ is the cross section for two images with a splitting angle
$ > \Delta\theta$.

According to the Press-Schechter theory, the comoving number density
with mass in the range $(M,M+dM)$ is given by
\begin{eqnarray}
     n(M,z)\,dM = {\rho_0\over M}\, f(M,z)\, dM\,. \label{eq:p-s0}
\end{eqnarray}
with the  matter density of universe today $\rho_0 = \Omega_M
\rho_{crit,0}$ and the critical matter density at present
$\rho_{{\rm crit},0}=3 H_0^2/(8\pi G)$.
 $f(M,z)$ is the Press-Schechter function, and we shall utilize the modified
 form by Sheth and Tormen (1999)
 \begin{eqnarray}
     & & f(M,z) = - {0.383\over \sqrt{\pi}}  {\delta_c\over \Delta^2} {d\Delta\over dM}
 \left[1+\left({\Delta^2\over0.707 \delta_c^2 }\right)^{0.3} \right] \times {\rm
 exp} \left[- {0.707\over2} \left({\delta_c\over
 \Delta}\right)^2\right]\,,
     \label{eq:p-s} \\
 & & \Delta^2 (M,z) = \int_0^{\infty} {dk\over k} \Delta_k(k,z) W^2(kr)
\end{eqnarray}
where $\Delta$ is the variance of the mass fluctuations(Eisenstein
and Hu 1999) and parameter $\delta_c (z)$ is the linear overdensity
threshold for a spherical collapse(Wang and Steinhardt 1998;
Weinberg and Kamionkowski 2002).

For different density profile of dark halo, the lensing cross
section $\sigma$ can be calculated out based on the lensing
equation. We shall use the combined mechanism of SIS and NFW model
to explain the whole experimental curve of strong gravitational
lensing statistic. For that a new model parameter $M_c$ was
introduced by Li and Ostriker (2002): lenses with mass $M<M_c$ have
the SIS profile, while lenses with mass $M > M_c$ have the NFW
profile. Then the differential probability is given by
 \[ dP/dM =
dP_{SIS}/dM\, \vartheta(M_c - M) + dP_{NFW}/dM\,\vartheta(M -
M_c)\] where $\vartheta$ is the step function, $\vartheta(x-y)=1$,
if $x>y$ and 0 otherwise. As the splitting angle $\Delta \theta$
is directly proportional to the mass $M$ of lens halos, the
contribution to large $\Delta \theta$ of SIS profile is depressed
by $M_c$. The lens data require a mass threshold $M_c \sim
10^{13}h^{-1}M_{\odot}$, which is consistent with the halo mass
whose cooling time equals the age of the universe today.

The likelihood function of the SGL splitting angle statistic is
defined as \beq {\rm L}(w) = (1-p(w))^{N-N_l}\prod_{i=1}^{N_l}
q_i(w). \label{eq:likh}\eeq $p(w)$ and $q_i(w)$ represent the
model-predicted lensing probabilities and the differential lensing
probabilities, respectively. They are related to $P$ in Equation
(\ref{eq:lp}) by an integration \begin{eqnarray}
   p(w) \equiv P_{\rm obs}(>\Delta\theta ) = \int\int B\, {d P(>\Delta\theta )\over
       dz}\, \varphi(z_s) dz dz_s\,,
    \label{eq:pobs}
\end{eqnarray}
and
\begin{eqnarray}
   q(w) \equiv {dP_{\rm obs}(>\Delta\theta ) \over d\Delta \theta } = \int\int B\, {d^2 P(>\Delta\theta )\over
       d\Delta\theta  dz}\, \varphi(z_s)  dz dz_s\,.
    \label{eq:pobs2}
\end{eqnarray}
${\rm B}$ is the magnification bias and can be found in our previous
work(Zhang et al. 2009). $\varphi(z_s)$ is the redshift distribution
of sources. Here we  take the Gaussian model by directly fitting the
redshift distribution of the subsample of CLASS statistical sample
provided by Marlow et al. (2000), which is given by (Zhang et al.
2009)
 \beq
 g(z_s) = {N_s \over \sqrt{2 \pi} \lambda } {\rm exp} \left[- {(z_s-a)^2}\over 2
 \lambda^2\right],
 \eeq with $N_s=1.6125; \; a=0.4224; \; \lambda=1.3761$.

\subsection{Joint Analysis and Numerical Results}

In this section, we will investigate the constraint on the
cosmological parameters from the joint analysis of (SNe + BAO) and
(SNe + BAO + SGL), respectively. For the two(or three) independent
observations, the likelihood function of a joint analysis is just
given by
 \beq
  L &=& L_{\rm SNe} \times L_{\rm BAO} \;\; (\times L_{\rm SGL})   \nonumber \\
    &=& \exp(-\chi_{\rm SNe}^2/2) \times \exp(-\chi_{\rm BAO}^2/2) \;\;  (\times L_{\rm SGL}).
 \eeq The statistic significance $\chi^2_{\rm BAO}$ and $\chi_{\rm SNe}^2$
can be obtained by using Equations (\ref{eq:baochi2}) and
(\ref{eq:snechi2}), respectively.  $L_{\rm SGL}$ is the likelihood
function of SGL statistic and can be obtained by using Equation
(\ref{eq:likh}).  For SGL data, we shall integrate the parameter $h$
from $0.4$ to $0.9$.

\begin{figure}
\epsscale{0.8} \plotone{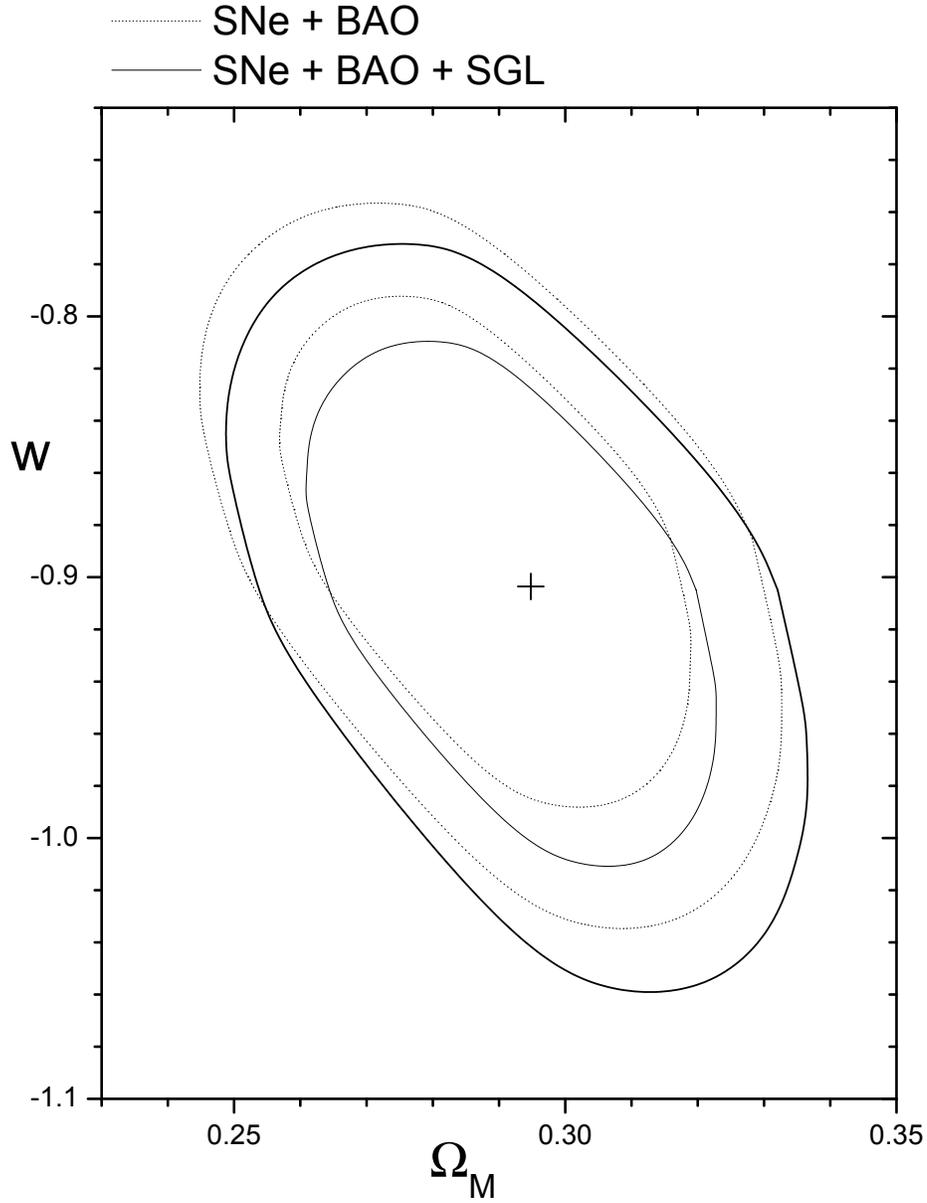}
\figcaption{\small  68\% C.L. and 95\% C.L. allowed regions of
($\Omega_M, w$) from the joint analysis of (SNe + BAO + SGL) (solid
lines) in comparison with the joint analysis of (SNe + BAO)(dotted
lines). The best-fit result from (SNe+ BAO + SGL) is $(\Omega_M, w)
= (0.29, -0.91)$. \label{fig:SNe6} }
\end{figure}

Let us first discuss the constraints for the constant $w$ case. In
figure \ref{fig:SNe6}, we show the constraints on $\Omega_M$ and the
constant $w$ from the joint analysis of (SNe + BAO + SGL). For a
comparison, the results of (SNe +BAO) have been shown as dotted
lines. The best fit result is $(\Omega_M, w) = (0.29, -0.91)$. The
$95\%$ C.L. allowed regions of constant $w$ and $\Omega_M$ are found
to be: $-1.06 \leq w \leq -0.77$ and $0.25 \leq \Omega_M \leq 0.34$.
Comparing with the results of (SNe + BAO) case, it is seen that the
results have only slight differences and the fitted $w$ is found to
be slightly smaller after adding the SGL data.

\begin{figure}
\epsscale{0.8} \plotone{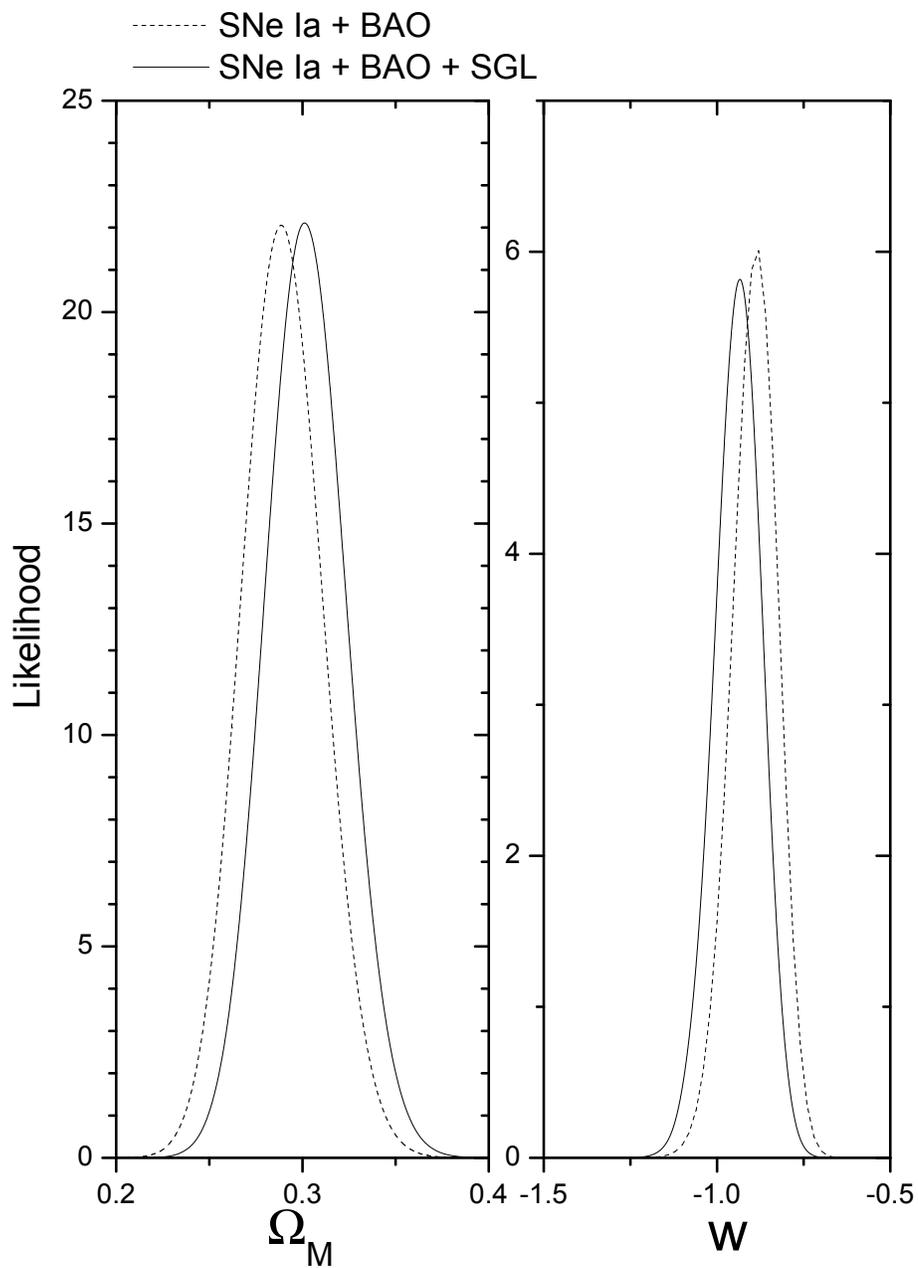}
\figcaption{\small   The likelihoods of the parameters $\Omega_M$
and  $w$ from the joint analysis of (SNe + BAO) and (SNe + BAO +
SGL), respectively. The maximum likelihood points are located at
$\Omega_M=0.29$ and $w=-0.88$for (SNe + BAO) and $\Omega_M=0.296$
and $w=-0.91$ for (SNe + BAO + SGL). \label{fig:SNe7} }
\end{figure}

Figure \ref{fig:SNe7} plots the likelihoods for the parameters
$\Omega_M$ and $w$ from the joint analysis of (SNe + BAO) and (SNe +
BAO + SGL), respectively. The maximum likelihood points are located
at $\Omega_M=0.29$ and $w=-0.88$ for (SNe + BAO) and
$\Omega_M=0.296$ and $w=-0.91$ for (SNe + BAO + SGL). It is
interesting to find that the parameters $w$ and $\Omega_M$ are
restricted to the range: $-1.17\leq w \leq -0.67$ and $0.23 \leq
\Omega_M \leq 0.37$.

\begin{figure}
\epsscale{0.8} \plotone{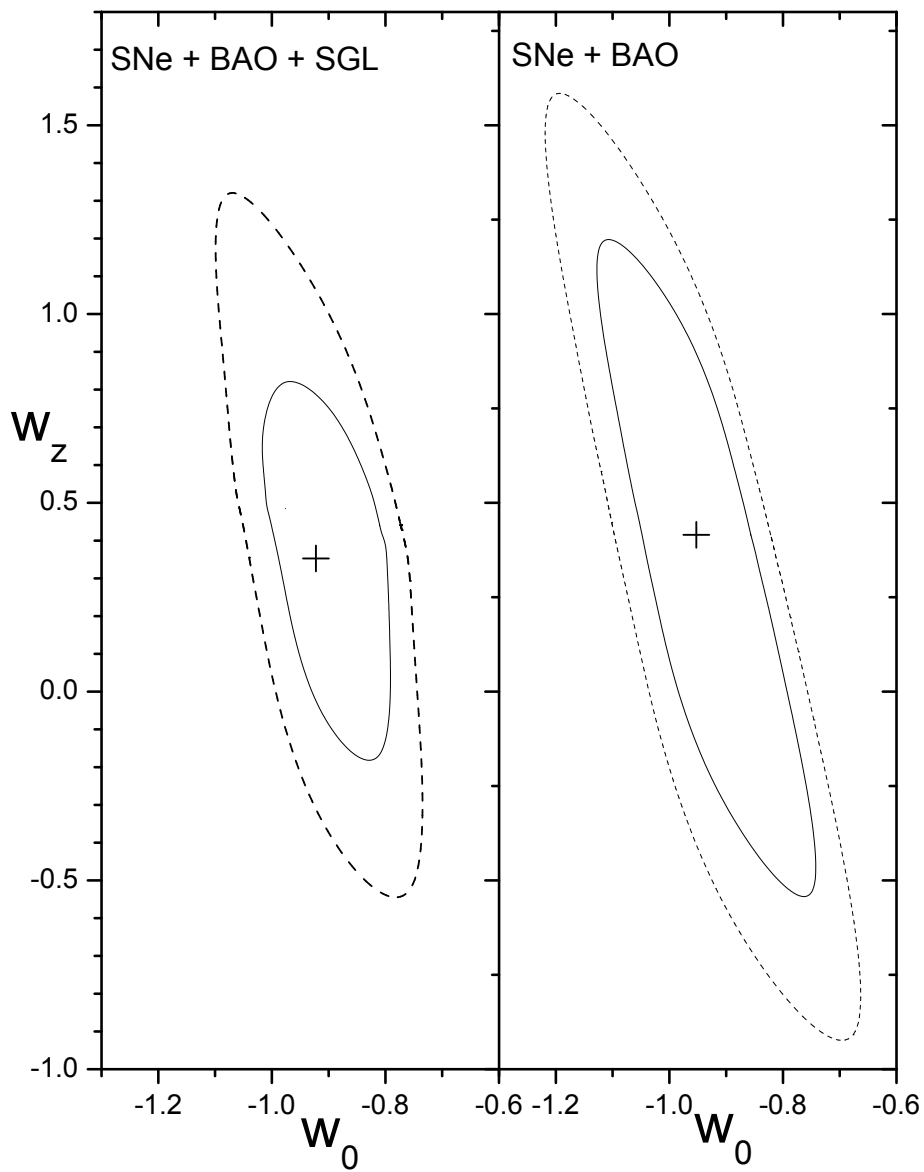}
\figcaption{\small the 68\% C.L. and 95\% C.L. allowed regions of
($w_0, w_z$) from the joint analysis of (SNe + BAO) and (SNe + BAO +
SGL), respectively. The crosshairs mark the best-fit point $(w_0,
w_z) = (-0.95, 0.41)$ for the (SNe + BAO) case and $(w_0,
w_z)=(-0.92, 0.35)$ for the (SNe + BAO + SGL) case.
\label{fig:SNe8}}
\end{figure}

After marginalizing the parameter $\Omega_M$, we obtain the
constraint on $(w_0, w_z)$ in figure \ref{fig:SNe8} from the joint
analysis of (SNe + BAO) and (SNe + BAO + SGL), respectively. The
crosshairs mark the best-fit point $(w_0, w_z) = (-0.95, 0.41)$ for
the (SNe + BAO) case and $(w_0, w_z)=(-0.92, 0.35)$ for the (SNe +
BAO + SGL) case. For the (SNe + BAO) case, the $95\%$ C.L. allowed
regions for the parameters $w_0$ and $w_z$ are found to be: $-1.22
\leq w_0 \leq -0.66$ and $-0.92 \leq w_z \leq 1.59$. For the (SNe +
BAO + SGL) case, the $95\%$ C.L. allowed regions for the parameters
$w_0$ and $w_z$ are found to be: $-1.10\leq w_0 \leq -0.72$ and
$-0.55\leq w_z \leq 1.32$. After adding the SGL data, the constraint
on the parameter $w_0$ is improved moderately, but for the parameter
$w_z$, the allowed region decreases by near half. The extra
constraint power on the time-varying $w(z)$ obtained through adding
SGL data is due to the larger redshift $0 < z < 3.0$ of the galaxies
in CLASS observational sample, in comparison with the reshift range
of SNe data $0 < z < 1.5$ and the redshift of BAO $z=0.35$. It can
be seen for the both cases that: (a) the most allowed region of
$w_z$ is above $w_z=0$; (b) in comparison with the cosmological
constant ($w_0, w_z) =(-1.0, 0.0)$, the joint analysis for both
cases favors more positive ($w_0, w_z$); (c) in comparison with the
results of SNe Ia data alone, the constraint on $w_z$ is much
improved and $w_0$ also gets better constrained.

\begin{figure}
\epsscale{0.8} \plotone{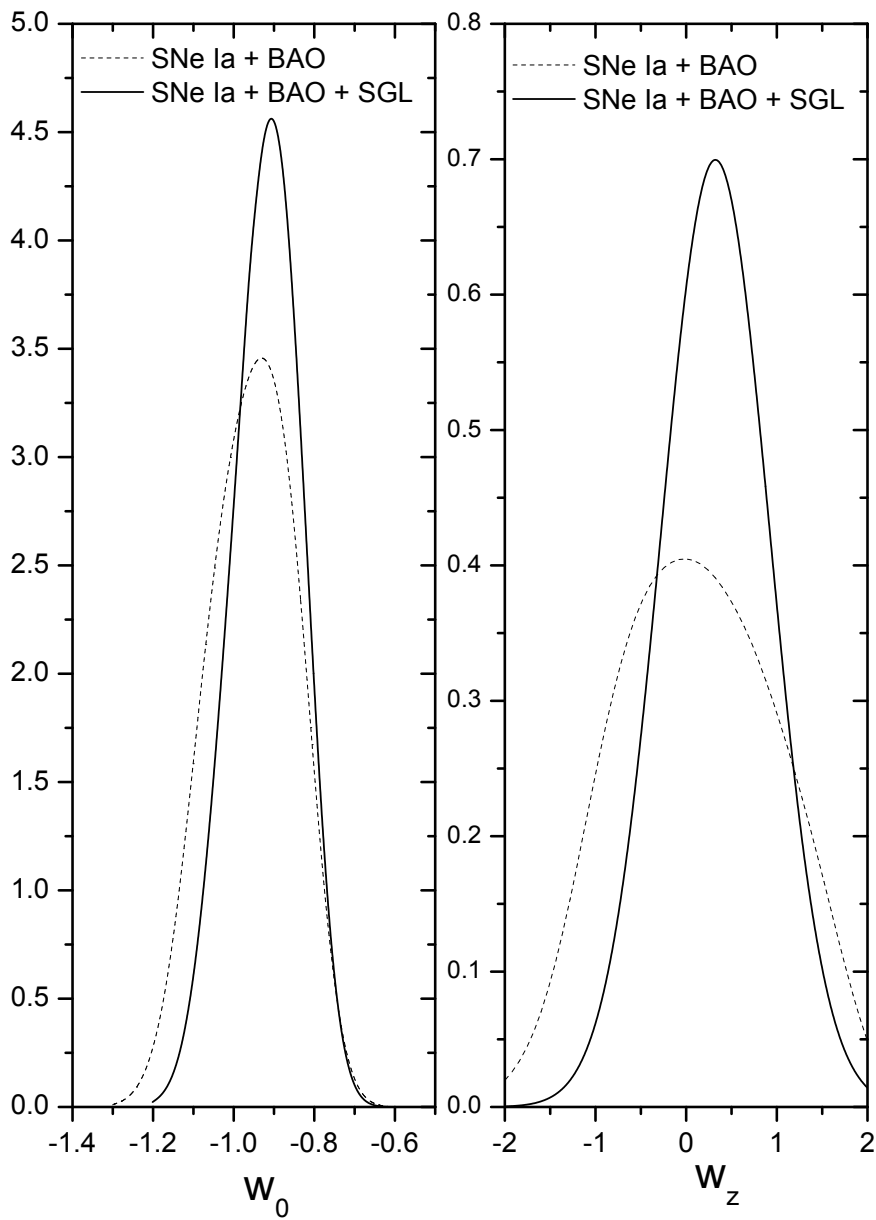}
\figcaption{\small the likelihoods of parameters $w_0$ and $w_z$
from the joint analysis of (SNe + BAO) and (SNe + BAO + SGL),
respectively. For (SNe + BAO) case, the maximum likelihood points
are located at $w_0=-0.94$ and $w_z = 0.0$, respectively. For (SNe +
BAO + SGL) case, the maximum likelihood points are $w_0=-0.91$ and
$w_z = 0.34$, repectively.\label{fig:SNe9}}
\end{figure}

Figure \ref{fig:SNe9} plots the likelihoods of parameters $w_0$ and
$w_z$ from the joint analysis of (SNe + BAO) and (SNe + BAO + SGL),
respectively. For (SNe + BAO) case, the maximum likelihood points
are located at $w_0=-0.94$ and $w_z = 0$. Note that $w_z=0$ implies
a constant equation-of-state of dark energy. For (SNe + BAO + SGL)
case, the maximum likelihood points are found to be $w_0=-0.91$ and
$w_z = 0.34$. We see that the parameters $w_0$ and $w_z$ are
restricted to be: $-1.20\leq w_0 \leq -0.67$ and $-1.0 \leq w_z \leq
2.0$.

\section{CONCLUSIONS}

We have carefully investigated, based on the latest SNe Ia data, BAO
and SGL statistic, the constraint on the equation-of-state parameter
$w$ of dark energy, especially for the time varying cases in the
flat universe.  The influences of the matter density $\Omega_M$ on
the fitting results  are carefully demonstrated. The typical
redshift measured by the three kinds of observations is $z \sim 1 $
and far smaller than the redshift of CMB involved, their constraints
on the parameter $w$ are effective and significant only for the
redshift region $z < 1.5$.

The influence of the equation-of-state parameter $w$ on the density
$\rho(z)$ of dark energy in the universe and the distance $d(z)$
makes SNe Ia data a powerful probe of dark energy. In this paper, we
have utilized the latest $324$ SNe Ia data provided by Hicken et al.
(2009) using MLCS17 light curve fitter with the best cuts $A_V \leq
0.5$ and $\Delta < 0.7$ to carefully investigate the constraint on
the equation-of-state parameter $w$ of dark energy.  For the
constant $w$ case, the best-fit results for the two correlated
parameters are found to be $(\Omega_M, w)=(0.358, -1.09)$. It is
seen that $\Omega_M$ is somewhat large in comparison with $\Omega_M
= 0.26$ of the concordance cosmology provided by WMAP five year
data(Komatsu et al. 2009); note that using a different
parameterization of dark energy, an alternative analysis (Huang et
al. 2009) presented a best-fitted result $\Omega_M=0.446$ from SNe
Ia data, which is even larger but still consistent with our result
at 95\% C.L. For the time-varying case,  after marginalizing the
parameter $\Omega_M$, we have obtained the fitting results $(w_0,
w_z)=(-0.73^{+0.23}_{-0.97}, 0.84^{+1.66}_{-10.34})$, which
indicates that (a) the SNe Ia data alone have only a poor constraint
power on the parameter $w_z$, an extra restriction of $\Omega_M$ is
necessary, so that the constraint of SNe Ia on the parameters $w_0$
and $w_z$ can be much improved; (b)the likelihood of parameter $w_z$
has a high non-Gaussian distribution.

The summary parameter of BAO can provide a standard ruler by which
the absolute distance of $z=0.35$ can be determined with $5\%$
accuracy. This ruler can be a $\Omega_M$ prior and has been used to
constrain dark energy. The strong gravitational lensing (SGL)
statistic is a useful probe of dark energy.  Through comparing the
observed number of lenses with the theoretical expected result, it
enables us to constrain the parameter $w$. We have used the latest
SNe Ia data together with  the BAO (and the CLASS statistical
sample) to constraint dark energy. For the constant $w$ case, the
results obtained from (SNe + BAO) and (SNe + BAO + SGL) only have a
slight difference. We have shown that: (a) for the (SNe + BAO) case,
the best fit results of the parameters $(\Omega_M, w)$ are $(0.287,
-0.885)$  and for the (SNe + BAO + SGL) case, the best fit point is
$(\Omega_M, w) = (0.298, -0.907)$; (b) the fitting results are found
to be $\Omega_M = 0.29^{+0.03}_{-0.03}$ and  $w =
-0.91^{+0.10}_{-0.10}$ for the (SNe + BAO + SGL), which are
consistent with the $\Lambda \rm CDM$ at 95\% C.L.; (c) the most
allowed region of parameter $w$ is above the line $w=-1$. Comparing
with the fitting results from the SNe Ia data alone, we have found:
(a) the allowed region at 95\% C.L. for $\Omega_M$ is reduced to
one-fifth; (b)  the best fit value of $w$ is almost not changed but
its variance is reduced very much.

For the time-varying case $w(z)$  after marginalizing $(\Omega_M)$,
we have obtained the fitting results $(w_0,
w_z)=(-0.95^{+0.45}_{-0.18}, 0.41^{+0.79}_{-0.96})$ for the (SNe +
BAO) case and $(w_0, w_z)=(-0.92^{+0.14}_{-0.10},
0.35^{+0.47}_{-0.54})$ for the (SNe + BAO + SGL) case. It has been
seen that the adding of the SGL data makes the constraints on
parameter ($w_0, w_z$) to be much improved. For both cases, the most
allowed region of $w_z$ is above $w_z=0$, which indicates that the
data from  the three observations (SNe + BAO + SGL) disfavor the
phantom type models. Comparing with the fitting results from the
latest SNe Ia data alone, we have observed that: (a) the best fit
values for $w_0$ are decreased by over $0.2$ and the variances are
approximately reduced to one-fourth; (b) the best fit values of
$w_z$ are decreased by $0.49$ and the variances are reduced to
one-twelfth.

In conclusion, the joint analysis of the latest MLCS17 data set
given by Hicken et al. (2009), summary parameters of BAO and SGL
data have provided an interesting constraint on the
equation-of-state parameter $w$ of dark energy, especially for the
time-varying case with parameters ($w_0, w_z$).  A large number of
SNe Ia samples with reduced systematical uncertainties in the near
future, together with possible new observations on BAO and SGL
statistic, would be very useful to understand the properties of dark
energy.

\section*{Acknowledgments}
The author (YLW) would like to thank Q.G. Huang and M. Li for useful
discussions. This work was supported in part by the National Basic Research Program of China (973 Program)
under Grants No. 2010CB833000, the National Science
Foundation of China (NSFC) under the grant \# 10821504, 10975170 and
the Project of Knowledge Innovation Program (PKIP) of Chinese
Academy of Science.


\newpage

\centerline{REFERENCES}

\begin{itemize}
\item[] 
Albrecht, A. et al., 2006, arXiv:astro-ph/0609591
\item[] 
Astier, P., et al. 2006, A\&A, 447, 31
\item[] 
Barger, V. Guarnaccia, E and Marfatia, D. 2006, Phys.Lett. B635
61-65
\item[] 
Biswas, R., and Wandelt, B. D., 2009, arXiv:0903.2532
\item[] 
Branch, D, 1998, Ann. Rev. Astron. Astrophys. 36, 17;
\item[] 
 Browne, I.A., et al. 2003, MNRAS, 341, 13
\item[] 
Chae, K.-H. 2007, ApJ, 658, 71
\item[] 
Chevallier, M., and Polarski, D.
Int. J. Mod. Phys. D10, 213 (2001)
\item[] 
Caldwell, R. R., Dave, R., and Steinhardt, P. J. 1998, Phys. Rev.
Lett., 80, 1582
\item[] 
Caldwell, R. R., Phys. Lett. B 545, 23(2002)
\item[] 
Cole, S. et al. 2005, MNRAS, 362, 505
\item[] 
Davis, T. M. et al. 2007, ApJ, 666, 716
\item[] 
Eisenstein, D. J., and Hu, W. 1999, ApJ, 511, 5
\item[] 
Eisenstein, D. J. et al., 2005, ApJ, 633, 560
\item[] 
Gibson, B.K., and Brook, C.B. 2001, arXiv:astro-ph/0011567
\item[] 
Guy, J., Astier, P., Nobili, S., Regnault, N., and Pain, R. 2005,
A\&A, 443, 781
\item[] 
Guy, J., et al. 2007, A\&A, 466, 11
\item[] 
Hicken, M. et al., 2009,  arXiv:0901.4804
\item[] 
Hinshaw, G.,  et al. 2009, ApJS, 180, 225
\item[] 
Huang, Q.G., Li, M., Li, X.D.,  and Wang, S. arXiv:0905.0797, 2009
\item[] 
G. Huetsi, A\&A, 2006 449, 891
\item[] 
Huterer, D. and Ma, C. P., 2004, ApJ, 600, 7
\item[] 
Jha, S., Riess, A. G., and Kirshner, R. P. 2007, ApJ, 659, 122
\item[] 
Kessler, R. et al. 2009, arXiv:0908.4274
\item[] 
Komatsu, E., et al. 2009, ApJS, 180, 330
\item[] 
Kowalski, M., et al. 2008, ApJ, 686, 749
\item[] 
Li, L. -X., and Ostriker, J.P. 2002, ApJ, 566, 652
\item[] 
Linder, E.V., Phys.Rev.Lett.90, 091301(2003)
\item[] 
Marlow, D. R., Rusin, D., Jackson, N., Wilkinson, P. N., and Browne,
I. W. A., 2000, AJ, 119, 2629
\item[] 
Padmanabhan, T. 2003, Phys. Rept. 380, 235
\item[] 
Percival W. J. et al., 2007, ApJ, 657, 51
\item[] 
Perlmutter, S. et al. 1999, ApJ, 517, 565
\item[] 
Porciani, C., and Madau, P. 2000, ApJ, 532, 679
\item[] 
Press, W.H., Teukolsky, S.A., Vetterling, W.T.,  and Flannery, B.P.,
1992,  Numerical Recipes in Fortran (New York: Cambridge University
Press)
\item[] 
Riess, A. G., et al. 1998b, AJ, 116, 1009
\item[] 
Riess, A. G., et al. 2004, ApJ, 607, 665
\item[] 
Riess, A. G., et al. 2007, ApJ, 659, 98
\item[] 
Riess, A. G., et al. 2009,  arXiv:0905.0697
\item[] 
Ratra, B., and Peebles, P. J. E. 1988, Phys. Rev. D, 37, 3406
\item[] 
Sandage, A., Saha, A., Tammann, G. A., Labhardt, L., Panagia, N.,
and Macchetto, F. D. 1996, ApJ, 460, L15
\item[] 
Sandage, A.,  Tammann, G.A., Saha, A., Reindl,  B.,  Macchetto,
F.D., and Panagia, N., 2006, ApJ, 653, 843
\item[] 
Sarbu, N., Rusin, D., and Ma, C.-P. 2001, ApJ, 561, L147
\item[] 
Schaefer, B.E., 1996, ApJ, 459, 438
\item[] 
Schneider, P., Ehlers, J., and Falco, E. E. 1992, Gravitational
Lenses (Berlin: Springer-Verlag)
\item[] 
Sheth R. K., and Tormen G., 1999, MNRAS, 308, 119
\item[] 
Wang, L., and Steinhardt, P. J. 1998, ApJ, 508, 483
\item[] 
Weinberg, N. N., and Kamionkowski, M. 2002, MNRAS, 337, 1269
\item[] 
Wood-Vasey, W. M. et al., 2007, ApJ, 666, 694w
\item[] 
York, D.G., et al., 2000, AJ, 120, 1579
\item[] 
Zhan, H. and  Knox, L.,  2006, arXiv:astro-ph/0611159
\item[] 
Zhang, Q.J., Cheng, L.M. and Wu, Y.L.,  2009, ApJ, 694, 1402

\end{itemize}

\clearpage

\end{document}